\documentclass{article}
\usepackage{spconf,amsmath,graphicx,amsmath,amsthm,paralist,amssymb,amsfonts}
\usepackage{tikz}
\usepackage{graphicx,comment}
\newtheorem{proposition}{Proposition}
\newtheorem{definition}{Definition}

\title{An ``augmentation-free'' rotation invariant classification scheme on point-cloud and its application to neuroimaging}
\name{Liu Yang and Rudrasis Chakraborty}
\address{University of California, Berkeley, USA}
\begin{document}
%
\maketitle
\begin{abstract}
Recent years have witnessed the emergence and increasing popularity of 3D medical imaging techniques with the development of 3D sensors and technology. However, achieving geometric invariance in the processing of 3D medical images is computationally expensive but nonetheless essential due to the presence of possible errors caused by rigid registration techniques. An alternative way to analyze medical imaging is by understanding the 3D shapes represented in terms of point-cloud. Though in the medical imaging community, 3D point-cloud processing is not a ``go-to'' choice, it is a canonical way to preserve rotation invariance. Unfortunately, due to the presence of discrete topology, one can not use the standard convolution operator on point-cloud. {\it To the best of our knowledge, the existing ways to do ``convolution'' can not preserve the rotation invariance without explicit data augmentation.} Therefore, we propose a rotation invariant convolution operator by inducing topology from hypersphere. Experimental validation has been performed on publicly available OASIS dataset in terms of classification accuracy between subjects with (without) dementia,  demonstrating the usefulness of our proposed method in terms of \begin{inparaenum}[\bfseries (a)] \item model complexity \item classification accuracy and last but most important \item invariance to rotations.\end{inparaenum}

\end{abstract}

\begin{keywords}
Rotation invariance, spherical convolution, attention, dementia
\end{keywords}
\section{Introduction} \label{intro}
Since the dawn of medical image analysis, researchers have been using 3D imaging to capture structure of the brain. Throughout the last decade, this community has seen the emergence of deep learning due to its power to capture the local structure. One of the major challenges to deal with medical images is the misalignment in the collection of data. Although this can be addressed by image registration, yet the registration techniques incur error which propagates to the subsequent analysis. A way to address this issue is by making the model invariant to the geometric transformations, which entails data augmentation with an increase in model complexity and training time. This leads to the voyage of exploring alternatives of 3D structure representation. Point-cloud is an efficient way to represent 3D structures \cite{qi2017pointnet,qi2017pointnet++} because of its important geometric properties. Due to the lack of a smooth topology, standard convolution can not be applied on point-cloud. One of the popular approaches to do point convolution \cite{zhou2018voxelnet} is to divide the point-cloud into voxels and then extract some features using 3D convolution. However, this method suffers from the possible sparsity of point-clouds which results in multiple empty voxels. One possible solution is to use multi-layer perceptron (MLP) to extract features from each point \cite{qi2017pointnet} or from a local neighborhood around each point \cite{qi2017pointnet++}. Unfortunately, all these methods are susceptible to random rotations which makes them incapable to efficiently deal with the error caused by rigid registration. In this work, we try to address this problem by developing an inherently rotation invariant model. 

\begin{figure}[!ht]
 \centering
\includegraphics[width=0.06\textwidth, height=1.6cm]{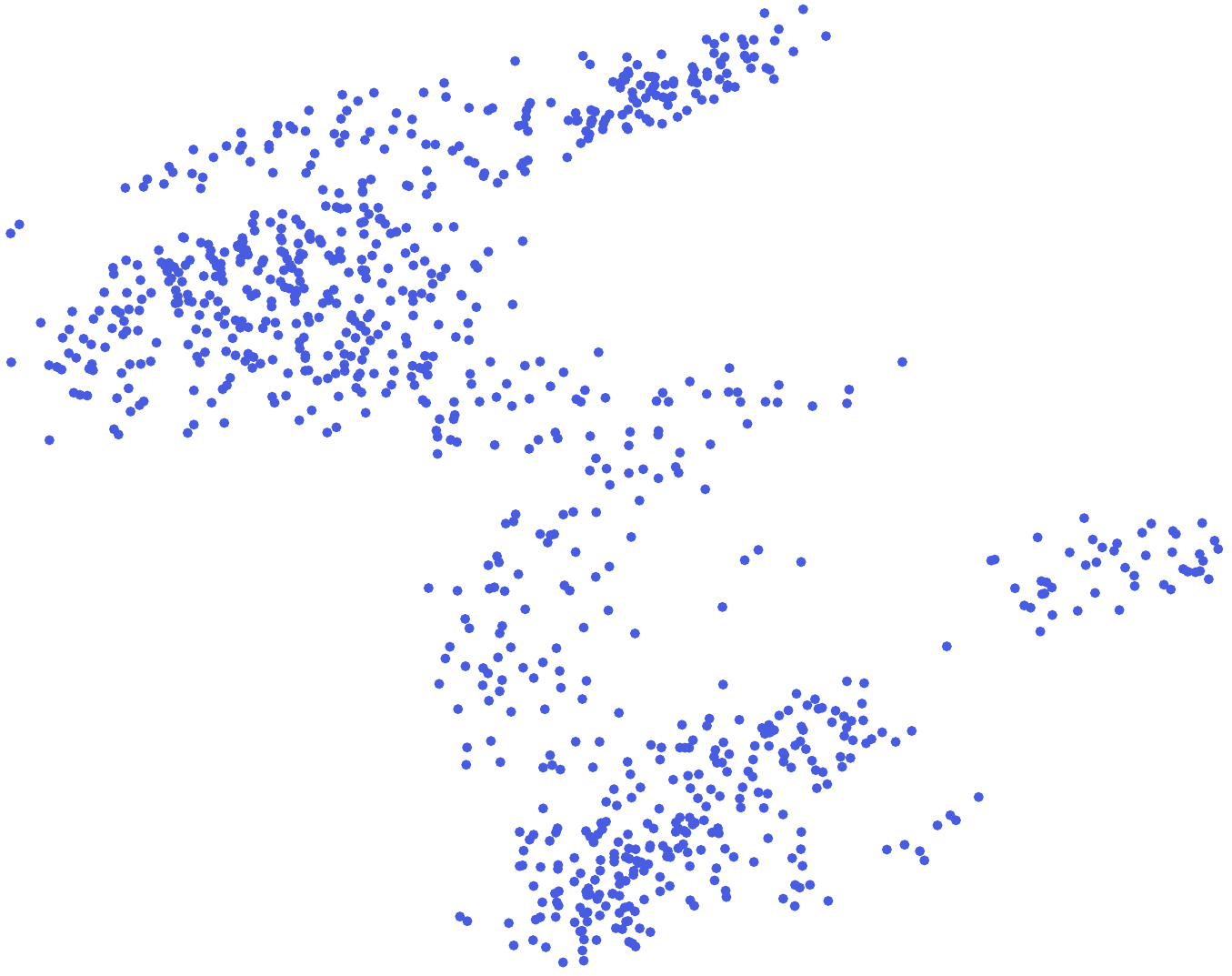}
 \includegraphics[width=0.28\textwidth, height = 1.6cm]{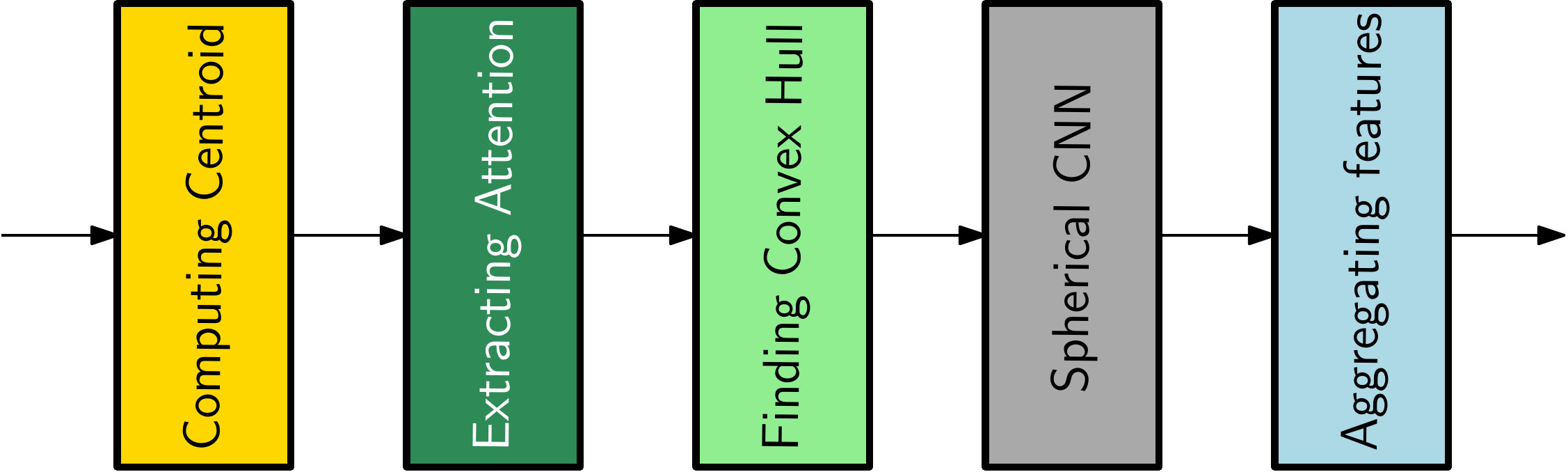}
\includegraphics[width=0.03\textwidth, height=1.6cm]{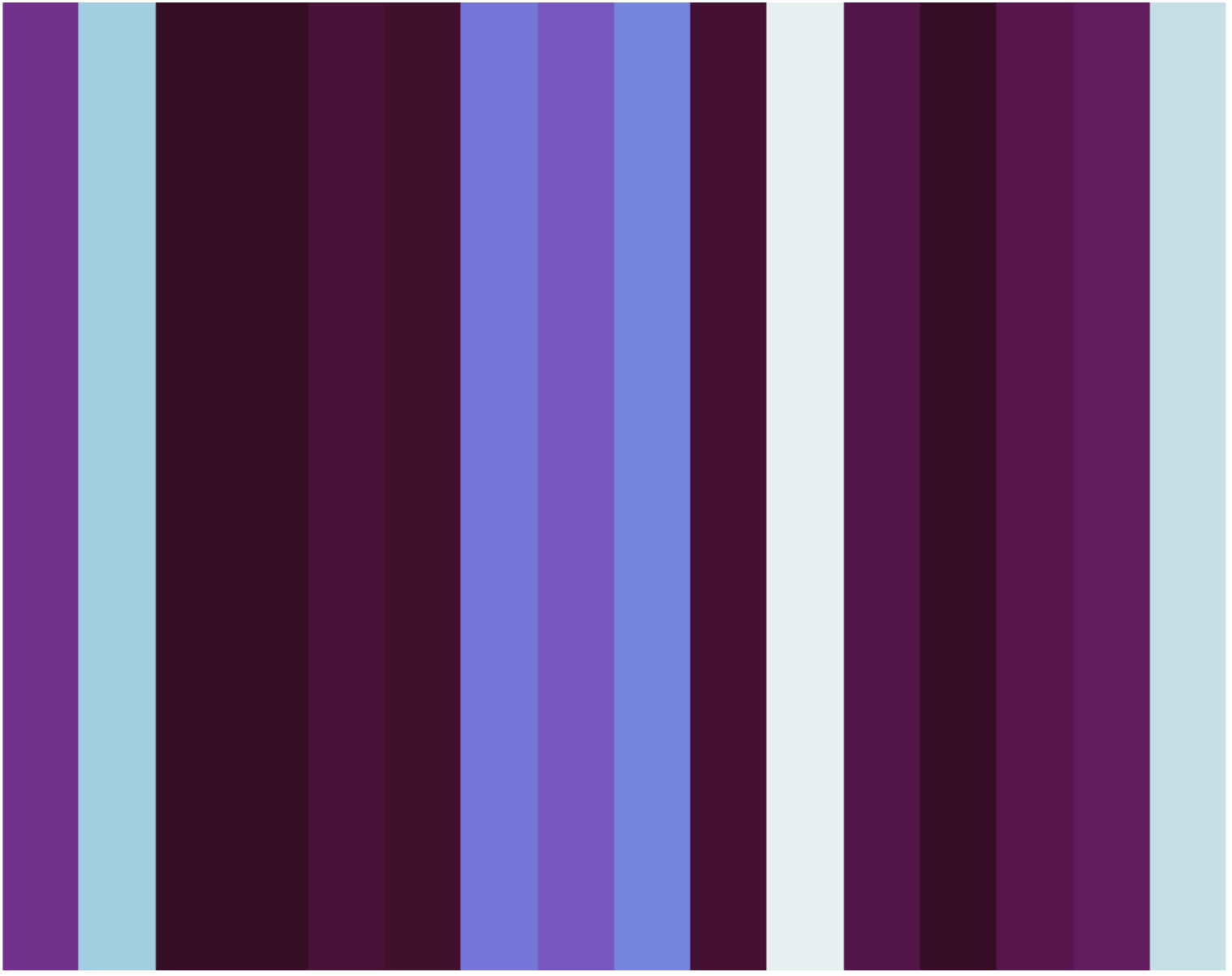}
\caption{Schematic of our proposed framework}
      \label{fig0}
\end{figure}

In recent years, several researchers \cite{chakraborty2017statistics,muralidharan2012sasaki} have proposed methods to do discrimination between subjects with and without dementia. Some of the popular approaches include using the 3D volume of the region of interest (ROI) and analyzing the shape of the anatomical structures of interest. In order to do that, the researchers either proposed techniques to map the 3D volume in high dimensional space \cite{chakraborty2017statistics} or mapped the shape of the anatomical structure on the complex projective space, i.e., Kendall's shape space \cite{muralidharan2012sasaki}. 

\begin{figure}[!ht]
 \centering
\includegraphics[scale=0.43]{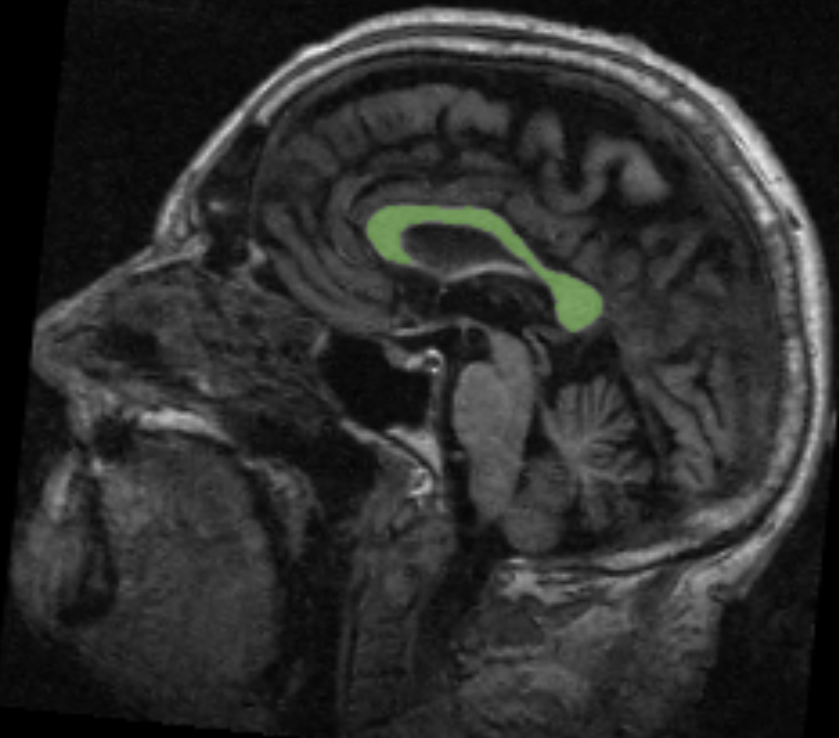}
\caption{Sample brain MR scan with corpus callosum mask overlayed}
      \label{fig-1}
\end{figure}

In this work, we use our proposed approach to discriminate between demented and non-demented subjects on publicly available OASIS dataset \cite{fotenos2005normative} based on the shape of the corpus callosum (as shown in Fig. \ref{fig-1}). It is well-accepted that thinning of corpus-callosum structure is related to dementia. We first segment the corpus callosum structure from the 3D image. Then we use point-cloud sampled from the the corpus callosum structure as the input of our proposed method. 

{\bf Motivation:} In medical imaging, the analysis is highly sensitive to the error caused by the rigid registration due to the presence of large rotations in collection of data. This dictates the necessity of a rotation invariant model for the medical imaging. However, achieving the invariance on 3D image scans is computationally expensive as it entails data augmentation. This motivates us to present 3D image scans with point-clouds. Nonetheless, due to the lack of local neighborhood structure, it is not possible to use standard convolution on the 3D point-cloud. Therefore, we define a rotation invariant convolution operator on point-cloud  with the topology induced from sphere.

Our proposed method first constructs sphere around each point and collect response from the point-cloud. The responses are collected based on the inner product between the grid points on the sphere and the points in the point-cloud. We use spherical convolution on the collected responses to extract rotation-invariant features. Finally, we aggregate the invariant response for the entire point-cloud and use this feature to do classification of demented and non-demented subjects. A schematic of our proposed method is shown in Fig. \ref{fig0}.

The salient features of our proposed method are: \begin{inparaenum}[\bfseries (1)] \item The previous methods to deal with point-clouds either use MLP to define ``convolution'' or use locally flat structure to define ``convolution''. {\it To the best of our knowledge, this is the first attempt to define a convolution operator on 3D point-cloud.} \item In order to achieve geometric invariance, the popular approach is to use data augmentation which naturally increases the training time and model complexity. Here, we propose an ``augmentation-free'' rotation invariant convolution neural network (CNN) for point-cloud. {\it To the best of our knowledge, this is the first attempt to model medical images using 3D point clouds.} Our model is much leaner because of the induced spherical topology on the point-cloud instead of mapping the point-cloud into a high dimensional space. \item We perform an experimental evaluation in terms of classifying subjects with and without dementia on publicly available OASIS dataset.  \end{inparaenum}

\section{Proposed algorithm} \label{theory}

In this work, we propose a rotation invariant CNN using an induced spherical topology on the point-cloud. Though the formulation described below can be applied on $\mathbf{R}^n$ for any $n>0$, in this work we have restricted ourselves to $n=3$. We divide our proposed algorithm into five key components.

{\bf Computing the centroid of a point-cloud:} Given the point-cloud $X = \left\{\mathbf{x}_i\right\}_{i=1}^N\subset \mathbf{R}^3$, we compute the centroid of the point-cloud to be the nearest point to the mean of $X$. Formally, let us denote the centroid of $X$ to be $\mathbf{m}$. Then, $\mathbf{m}$ is defined as
\begin{align*}
\mathbf{m} = \Pi_{X}\left(\frac{1}{n} \sum_{i=1}^n \mathbf{x}_i\right), 
\end{align*}
where, $\Pi_{X}(\mathbf{x})$ is the projection of $\mathbf{x}$ in the set $X$. 

{\bf Extracting the ``attention'' from a point-cloud:} Given the point-cloud $X$, we extract the region of interest, i.e., ``attention'' to be a subset $Y \subset X$ as follows: \begin{inparaenum}[\bfseries (a)] \item Compute the directional part of the vector from $\mathbf{m}$ to each point, $\mathbf{x}_i$. Let the vector be denoted by $\mathbf{v}_i$. \item Pass the vector $\mathbf{v}_i$ through a fully-connected layer to get the confidence, $c_i \in [0, \infty)$, for selecting $\mathbf{x}_i$. \item Define a random variable following multinomial distribution with $\mathbf{c}$ as the parameter. \item Draw samples from this random variable to generate $Y$.\end{inparaenum} We call this subset $Y$ to be our region of interest. 

{\bf Finding Convex hull:} We extract convex hull using the following scheme as it is useful to capture the global structure of a geometric shape. At each point $\mathbf{y}_i \in Y \subset X$, we center a sphere with radius $r$, and collect the response from the set $Y$. This gives at each point $\mathbf{y}_i \in Y$, a function  $f_i: \mathbf{S}^2_r(\mathbf{y}_i) \rightarrow \mathbf{R}$. Now, given $\mathbf{z} \in \mathbf{S}^2_r(\mathbf{y}_i)$ (centered at $\mathbf{y}_i \in Y)$, we compute the response $f_i(\mathbf{z}) = \sum_{\mathbf{y} \not\in \mathbf{B}^2_r(\mathbf{y}_i)}\mathbf{z}^t(\mathbf{y} - \mathbf{y}_i)$, where $\mathbf{B}^2_r$ is the unit ball with radius $r$ centered at $\mathbf{B}^2_r(\mathbf{y}_i)$. 
\begin{proposition}
If the point-cloud $X$ is rotated by the matrix $R$, then the corresponding responses $\left\{f_i\right\}$ are also rotated by the matrix $R$.
\end{proposition}
This representation can be viewed as {\it putting omni-directional camera at each point and collecting the responses in each viewing direction.} This analogy makes one wonder: {\it Is there a necessity for $N$ cameras where $N$ is the number of points inside the region of attention?} Obviously, for a dense point-cloud the answer is no and hence we have incorporated a downsampling strategy as follows: \begin{inparaenum}[\bfseries{(a)}]  \item on each sphere centered at $\mathbf{y}_i$, collect the omni-directional responses from $Y$. \item choose top $\eta\%$ spheres with the largest responses. Let the chosen index set be $I \subset \left\{1, 2, \cdots, N\right\}$. \end{inparaenum} 
\begin{proposition}
The selected point set $\left\{\mathbf{y}_j\right\}_{j\in I}$ lie on the convex hull of the point-cloud. 
\end{proposition}

{\bf Extracting features with spherical Convolution:} After constructing sphere at $\left\{\mathbf{y}_j\right\}_{j \in I}$, we use spherical convolution to extract rotation equivariant features. We define spherical convolution as follows:
\begin{definition}[$\mathbf{S}^2$ convolution]
Given $f:\mathbf{S}^2 \rightarrow \mathbf{R}$ (the signal) and $w:\mathbf{S}^2 \rightarrow \mathbf{R}$ (the learnable kernel), we define the convolution operator $f\star w: \textsf{SO}(3) \rightarrow \mathbf{R}$ as $
\left(f\star w\right)(g) = \int_{\mathbf{S}^2} f(\mathbf{x}) w(g^{-1}\cdot \mathbf{x}) \omega(\mathbf{x}).
$
Here, $\omega$ is the volume density on $\mathbf{S}^2$. As showed in \cite{cohen2017convolutional,chakraborty2018h}, the above definition of spherical convolution is equivariant to the action of $\textsf{SO}(3)$. 
\end{definition}
As the output of spherical convolution is a function from $\textsf{SO}(3)$ to $\mathbf{R}$, we need $\textsf{SO}(3)$ convolution in order to develop a deep CNN architecture. We use the standard group convolution with respect to the Haar measure, which we define as follows.
\begin{definition}[$\textsf{SO}(3)$ convolution]
Given $f:\textsf{SO}(3) \rightarrow \mathbf{R}$ (the signal) and $w:\textsf{SO}(3) \rightarrow \mathbf{R}$ (the learnable kernel), we define the convolution operator $f\star w: \textsf{SO}(3) \rightarrow \mathbf{R}$ as $
\left(f\star w\right)(g) = \int_{\textsf{SO}(3)} f(h) w(g^{-1}h) \omega(h).
$
Here, $\omega$ is the volume density on $\textsf{SO}(3)$  with respect to the Haar measure. As obvious from the definition of group convolution this is equivariant to the action of $\textsf{SO}(3)$. 
\end{definition}  

We extract the rotation equivariant features by using $\mathbf{S}^2$ convolution and multiple $\textsf{SO}(3)$ convolutions. In between convolutions, we use ReLU and batchnorm operations. After these convolution layers, the output is equivariant to rotations, i.e.,  if $\mathbf{x} \mapsto \mathfrak{F}(\mathbf{x})$, then $g\cdot\mathbf{x} \mapsto g\cdot\mathfrak{F}(\mathbf{x})$ where, $g\cdot\mathbf{x}$ is the rotation of $\mathbf{x}$ using $g$.

We use an invariant layer after the convolution layers to compute the integrated response over $\textsf{SO}(3)$. The purpose of this invariant layer is to make the entire network invariant to rotations. 

\begin{figure*}[!ht]
 \centering
\begin{tikzpicture}
\node[inner sep=0pt] (fig1) at (0,0)
{\includegraphics[scale=0.2]{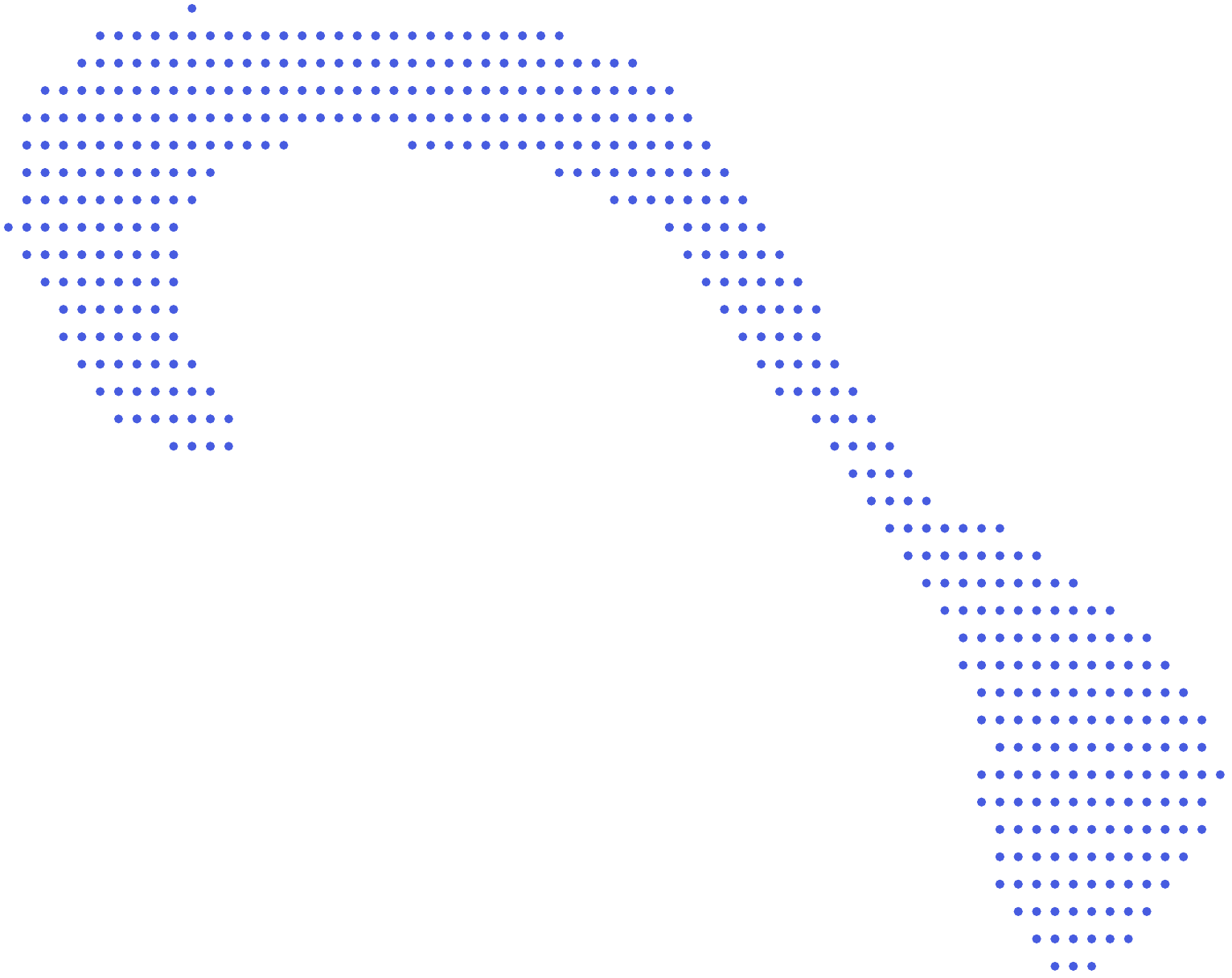}};
\node[inner sep=0pt] (fig2) at (4,0)
 {\includegraphics[scale=0.30]{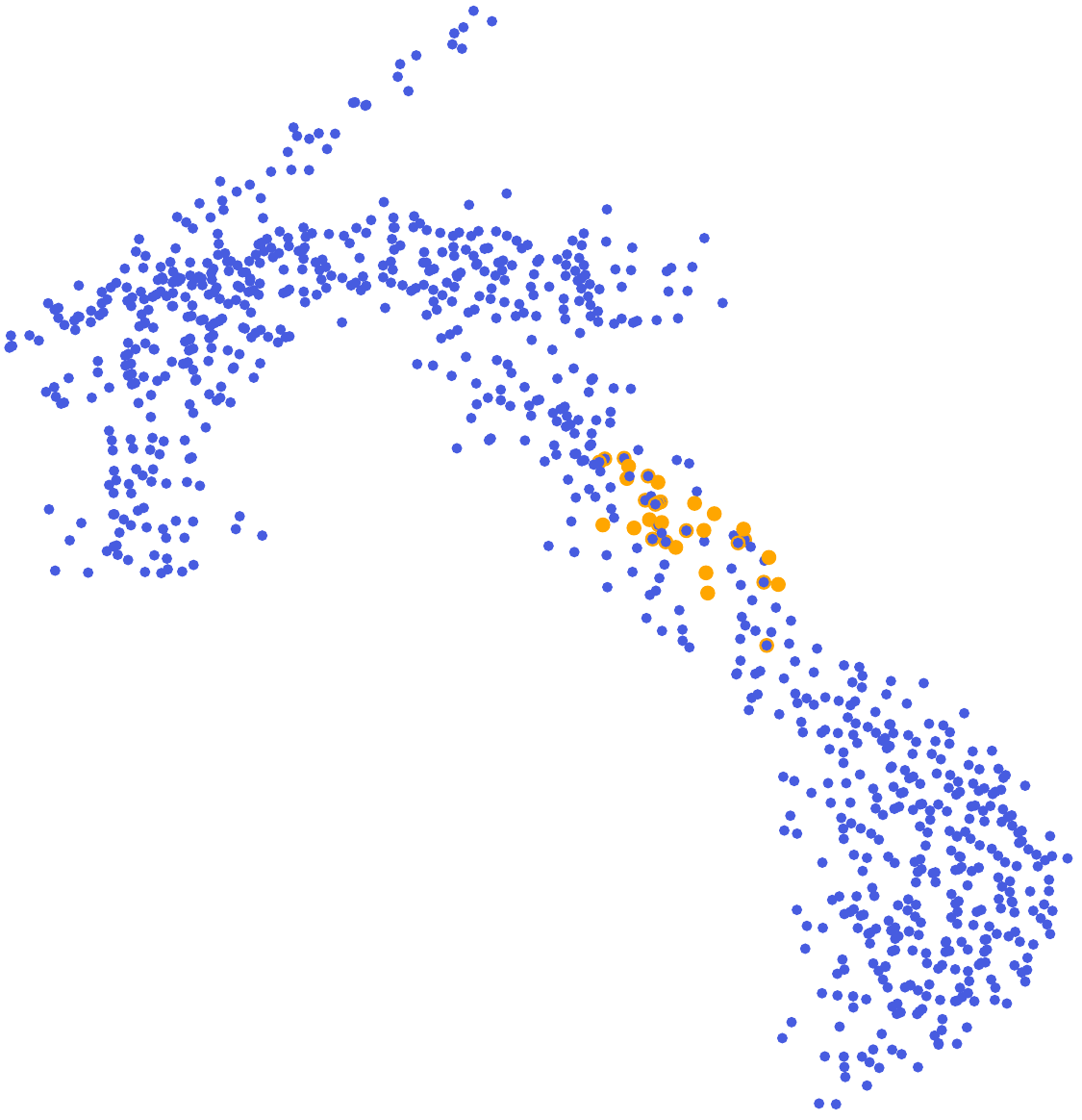}};
 \node[inner sep=0pt] (fig2) at (8,0)
{\includegraphics[scale=0.2]{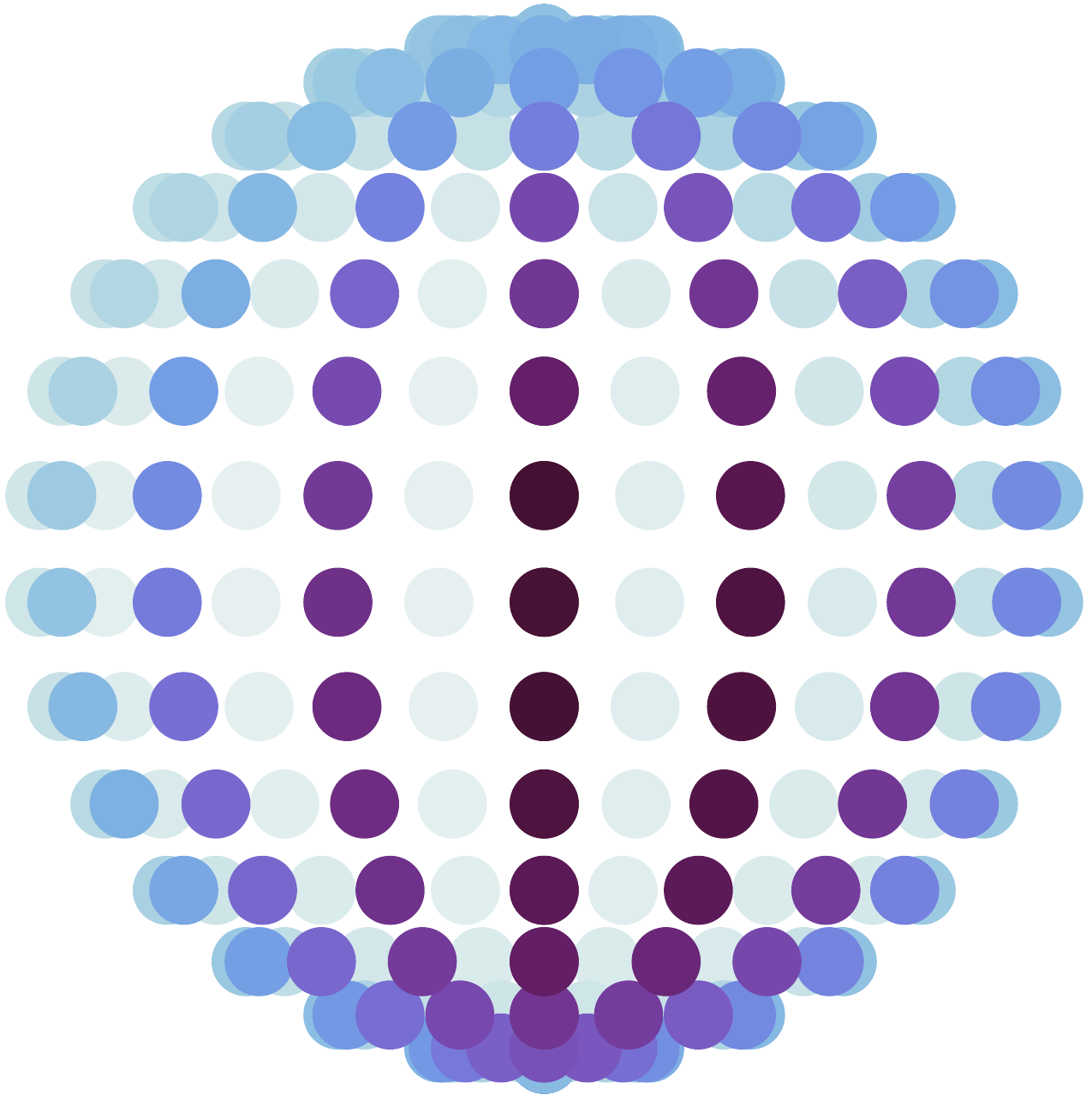}};
\node[inner sep=0pt] (fig2) at (11.5,0)
{\includegraphics[scale=0.2]{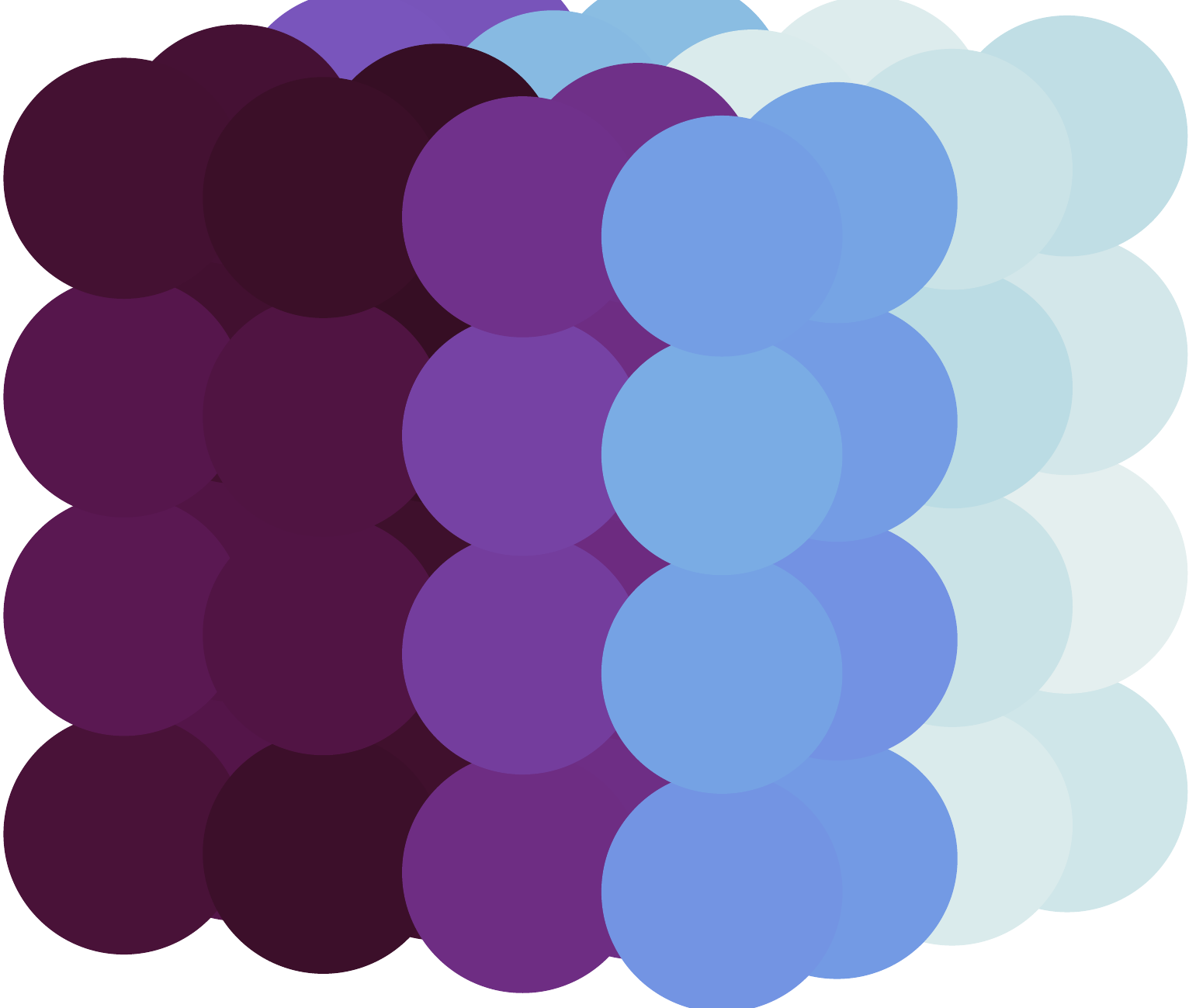}};
\node[inner sep=0pt] (fig2) at (15,0)
{\includegraphics[width=0.09\textwidth, height=2.5cm]{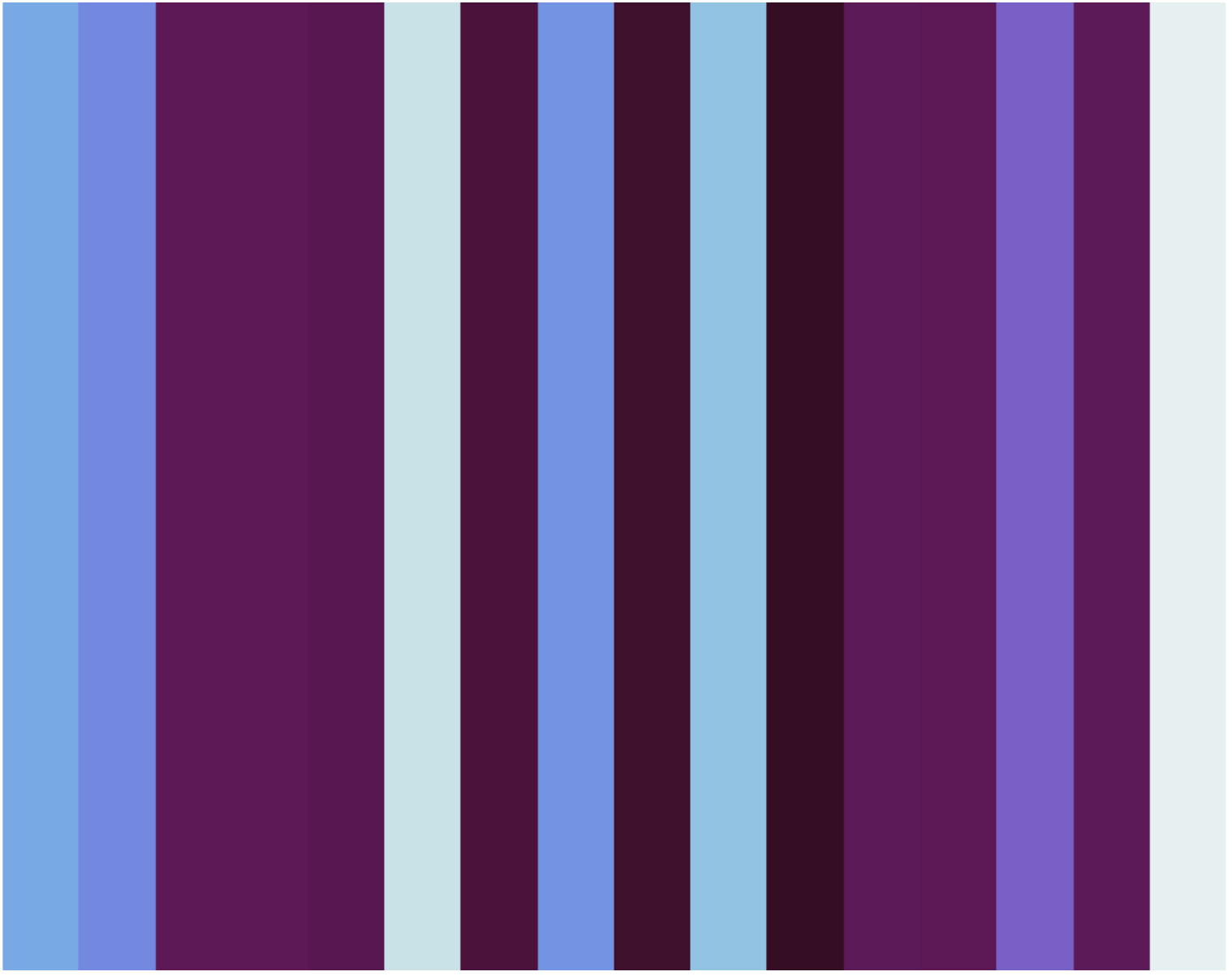}};
\end{tikzpicture}
\caption{The overview of our proposed model {\bf (a)} sample point-cloud {\bf (b)} point-cloud with attention {\bf (c)} response on $\mathbf{S}^2$ {\bf (d)} $\textsf{SO}(3)$ convolution response {\bf (e)} integrated response. }
      \label{fig3}
\end{figure*}

{\bf Aggregating features:} We extract rotation invariant spherical features for each point in the convex hull. With the purpose of classifying the point-cloud, we do aggregation of this invariant features collected over the convex hull. Given $\left\{\mathbf{s}_j\right\}_{j\in I}$, as the extracted features, we use global maxpooling over the convex hull. Thus, for each point-cloud $X$, we identify with a feature $\mathbf{s}_X = \max_{j \in I} \left\{\mathbf{s}_j\right\}$. We then use single fully connected layer to classify. An overview of our proposed method is given in Fig. \ref{fig3}.

In the next section, we will give the data description and the experimental details.

\section{Experimental results} \label{results}

This section consists of the data description followed by the details of experimental validation. 
\ \\
{\bf Data description:} In this section, we use OASIS data \cite{fotenos2005normative} to address the
classification of demented vs. non-demented subjects using
our proposed framework. This dataset contains at least two MR brain
scans of $150$ subjects, aged between $60$ to $96$ years old. For each
patient, scans are separated by at least one year. The dataset
contains patients of both sexes. In order to avoid gender effects, we take MR scans of male patients alone from three visits, which
resulted in the dataset containing $69$ MR scans of $11$ subjects with
dementia and $12$ subjects without dementia. This gives $33$ scans for subjects with dementia and $36$ scans for subjects without dementia. We first compute an atlas
(using the method in \cite{avants2009advanced}) from the $36 (=12
\times 3)$ MR scans of patients without dementia.

After rigidly registering each MR scans to the atlas, we segment out the corpus callosum region from each scan. We represent the shape of the corpus callosum as a 3D point-cloud. 
\ \\
{\bf Experimental details:} Below, we provide the experimental details involving choice of hyperparameters, data augmentation scheme and ablation study.

\begin{figure}[!ht]
 \centering
\vspace*{-1em}
\begin{tikzpicture}
\node[inner sep=0pt] (fig1) at (0,0)
    {\includegraphics[width=0.12\textwidth, height=2.3cm]{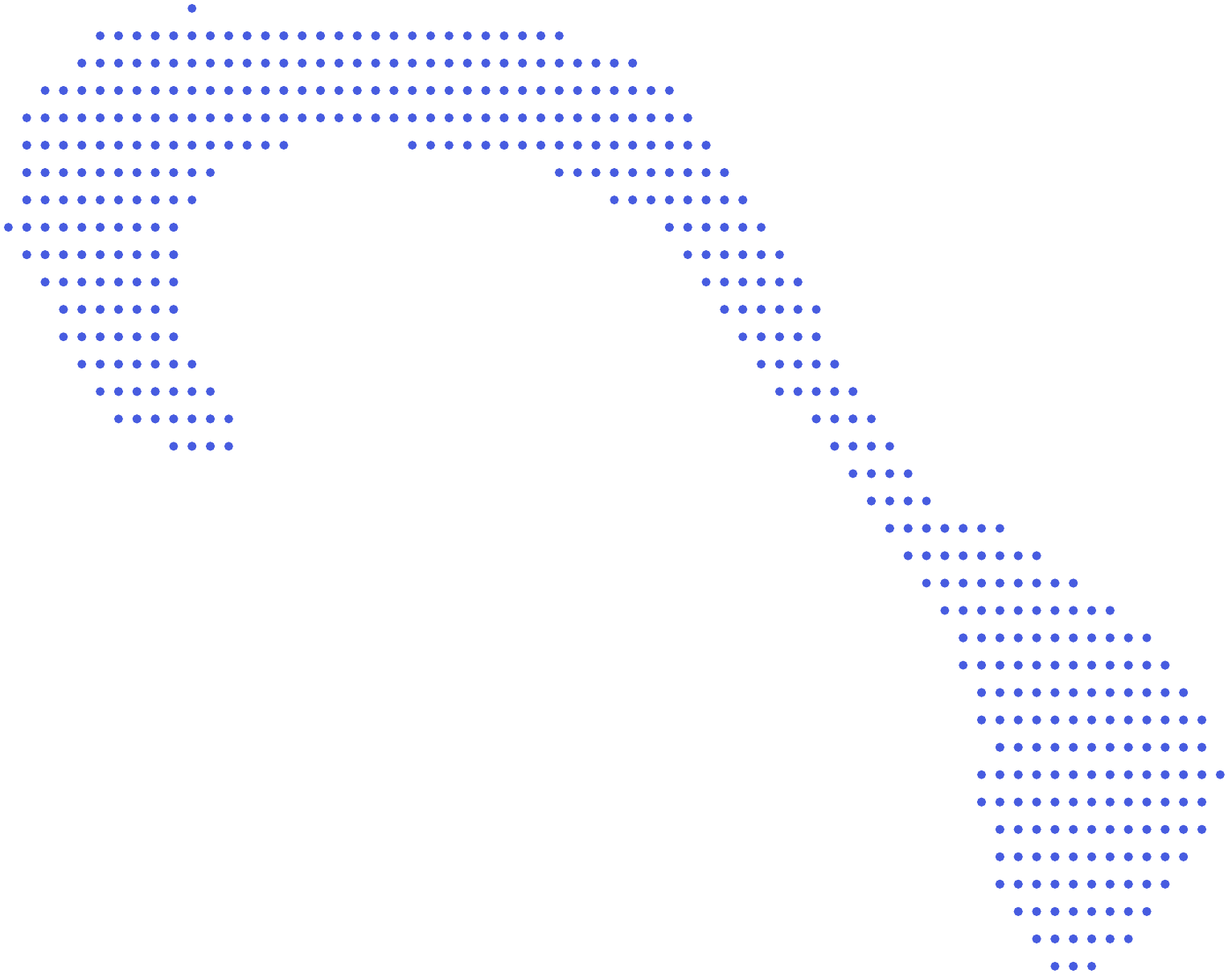}};
\node[inner sep=0pt] (fig2) at (4,0)
    {\includegraphics[width=0.12\textwidth, height=2.3cm]{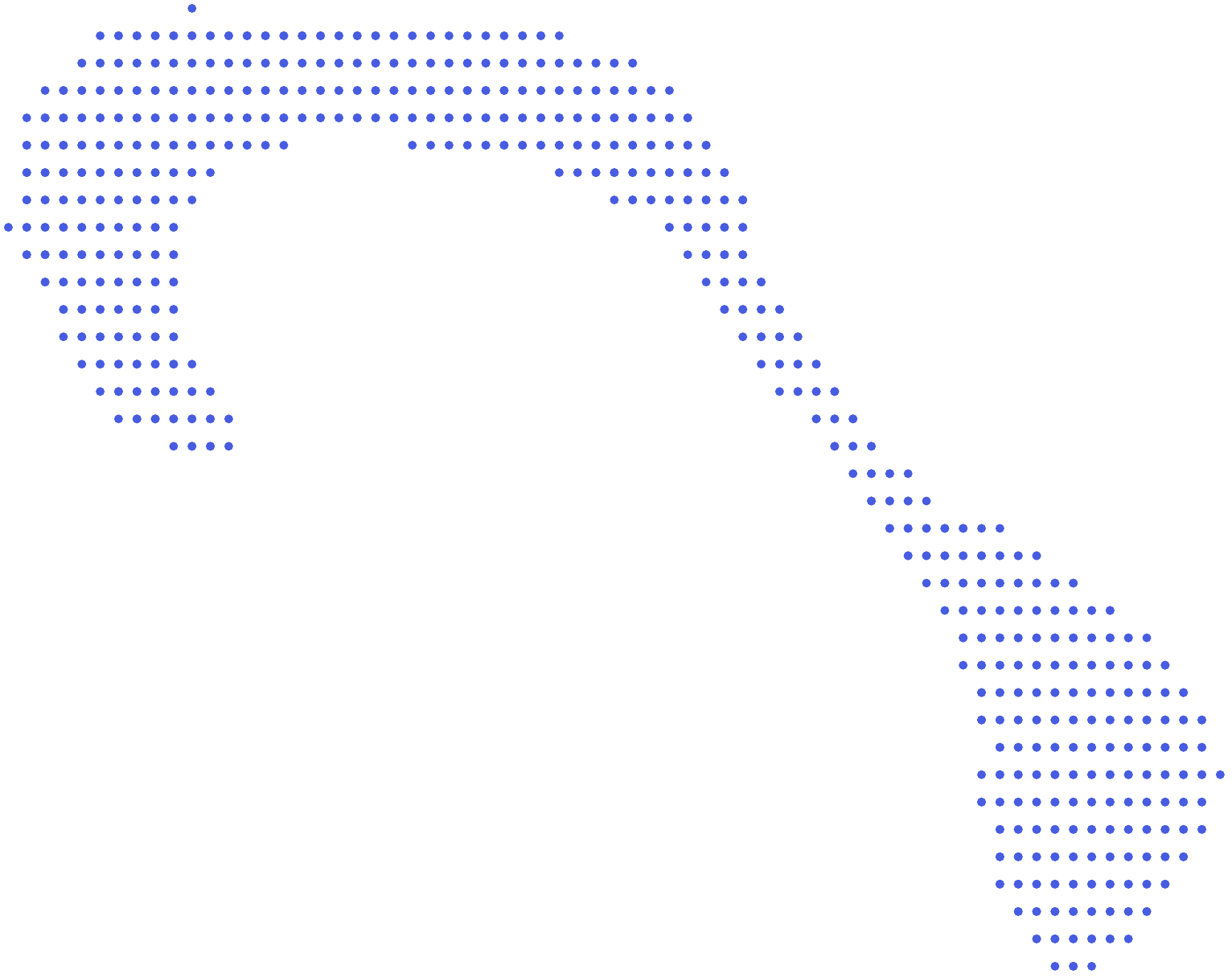}};
\end{tikzpicture}
\begin{tikzpicture}
\node[inner sep=0pt] (fig1) at (0,0)
    {\includegraphics[width=0.19\textwidth, height=2.9cm]{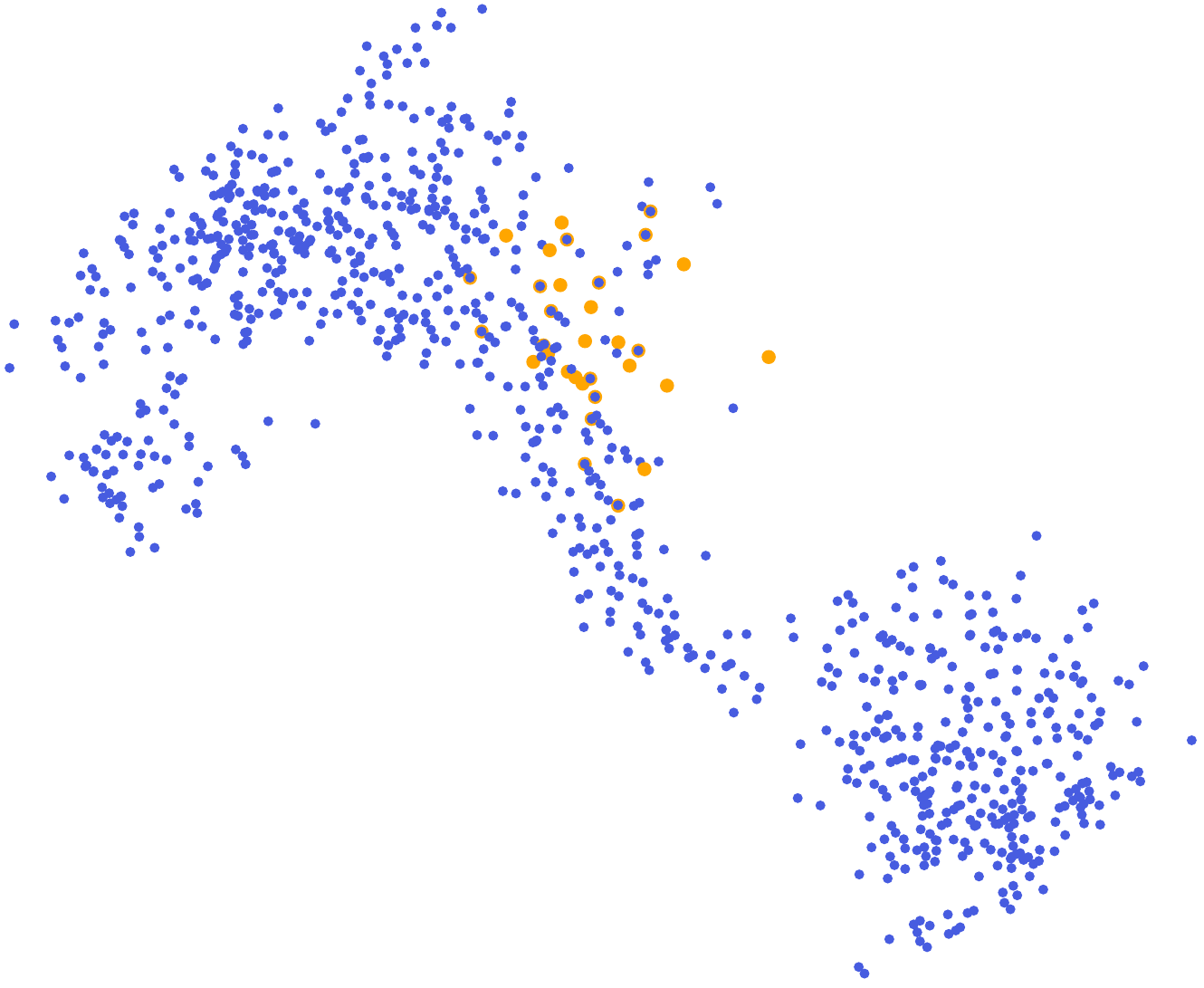}};
\node[inner sep=0pt] (fig2) at (4,0)
    {\includegraphics[width=0.17\textwidth, height=2.7cm]{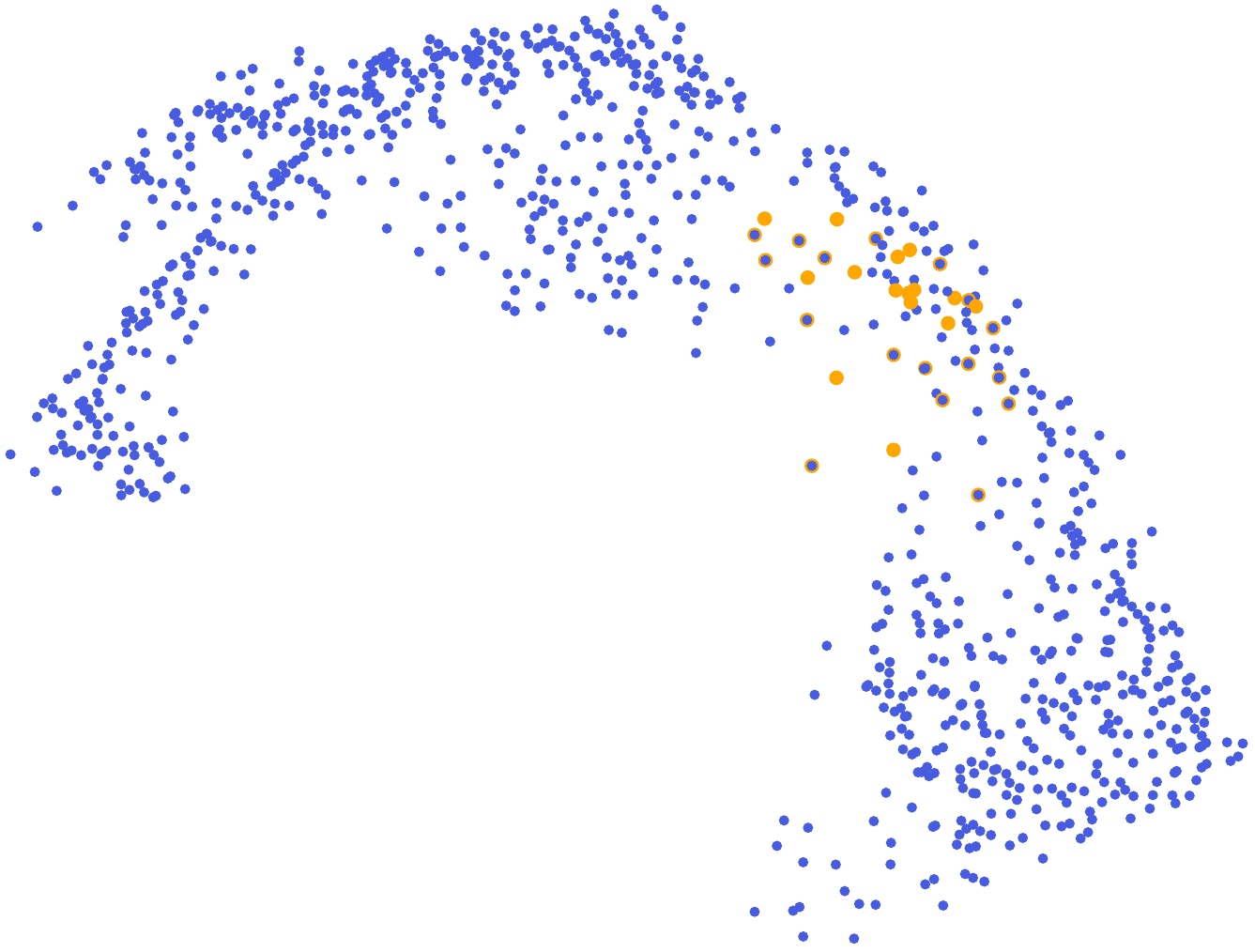}};
\end{tikzpicture}
\begin{tikzpicture}
\node[inner sep=0pt] (fig1) at (0,0)
    {\includegraphics[width=0.17\textwidth, height=2.7cm]{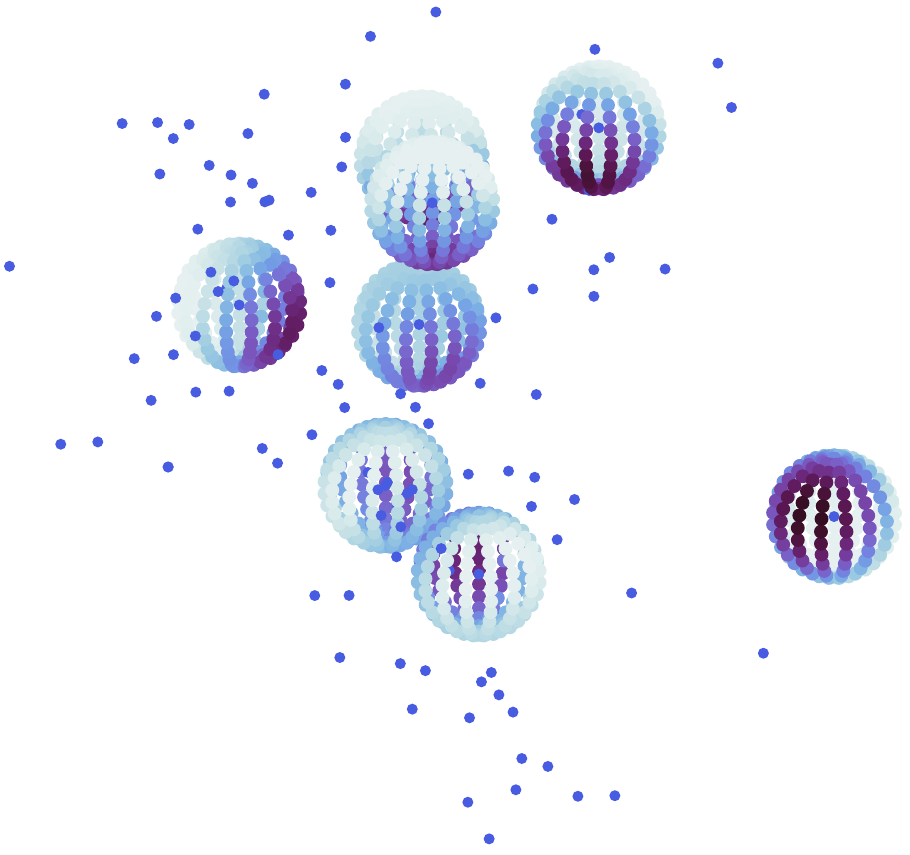}};
\node[inner sep=0pt] (fig2) at (4,0)
    {\includegraphics[width=0.12\textwidth, height=2.3cm]{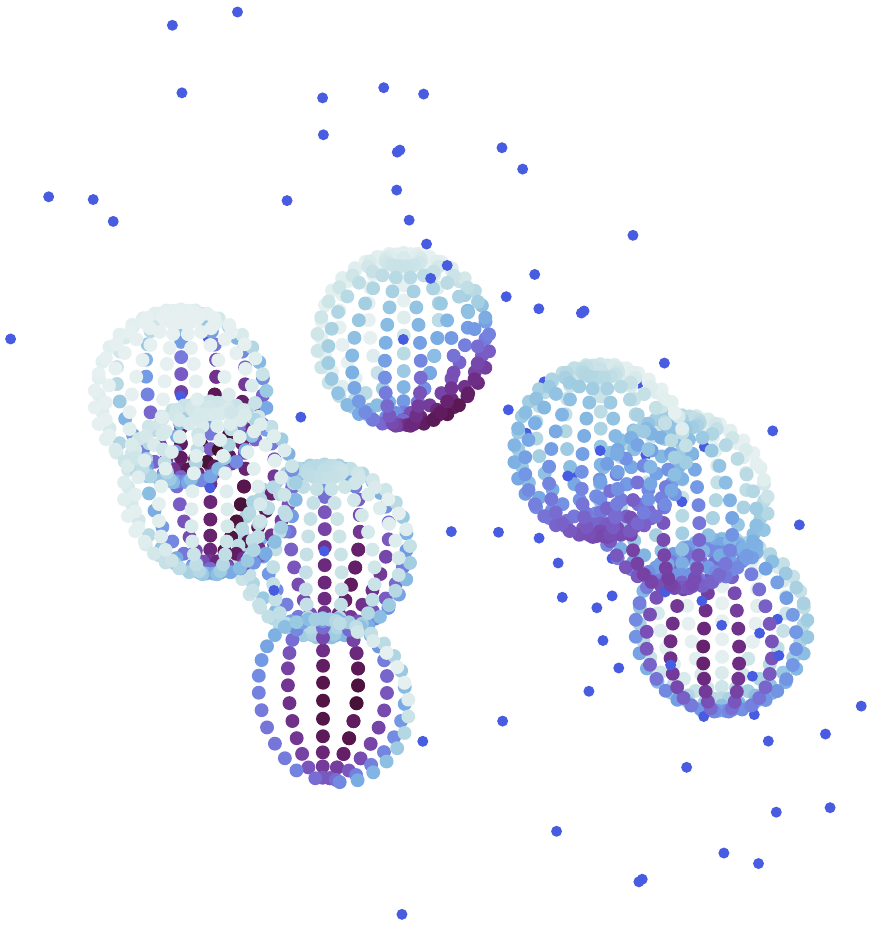}};
\end{tikzpicture}
\vspace*{-1em}
\caption{{\it (Top):} Sample point-cloud {\it (Middle:)} with attention region marked with ``orange'' {\it (Bottom:)} putting sphere around the convex hull points. (The first and second columns represent samples with non-dementia and dementia respectively.)}
      \label{fig1}
\end{figure}

{\bf Parameter selection:} The dataset consists of $928$ points in each point-cloud. We select $8$ points on the convex hull sampled from the region of interest. We use Adam optimizer with batch size $4$ and learning rate $0.005$. We use a spherical CNN model with one $\mathbf{S}^2$ and one $\textsf{SO}(3)$ convolution layer where input bandwidths are $4$ and $2$ and output bandwidths are $2$ and $1$ respectively. The number of output channels used are $8$ and $16$ respectively. {\it The total number of parameters used in this model is $2201$}. In Fig. \ref{fig1}, we show a representative point-cloud from demented and non-demented subjects with the corresponding attention area and the convex hull points. 

\begin{figure}[!ht]
 \centering
\begin{tikzpicture}
\node[inner sep=0pt] (fig1) at (0,0)
    {\includegraphics[width=0.12\textwidth, height=2.3cm]{non_rotate_whole.pdf}};
\node[inner sep=0pt] (fig2) at (6,0)
    {\includegraphics[width=0.12\textwidth, height=2.3cm]{non_rotate_response.pdf}};
    \draw[->, line width=0.2mm] (2.3,0) -- (3.9, 0);
\end{tikzpicture}
\begin{tikzpicture}
\node[inner sep=0pt] (fig1) at (0,0)
    {\includegraphics[width=0.12\textwidth, height=2.3cm]{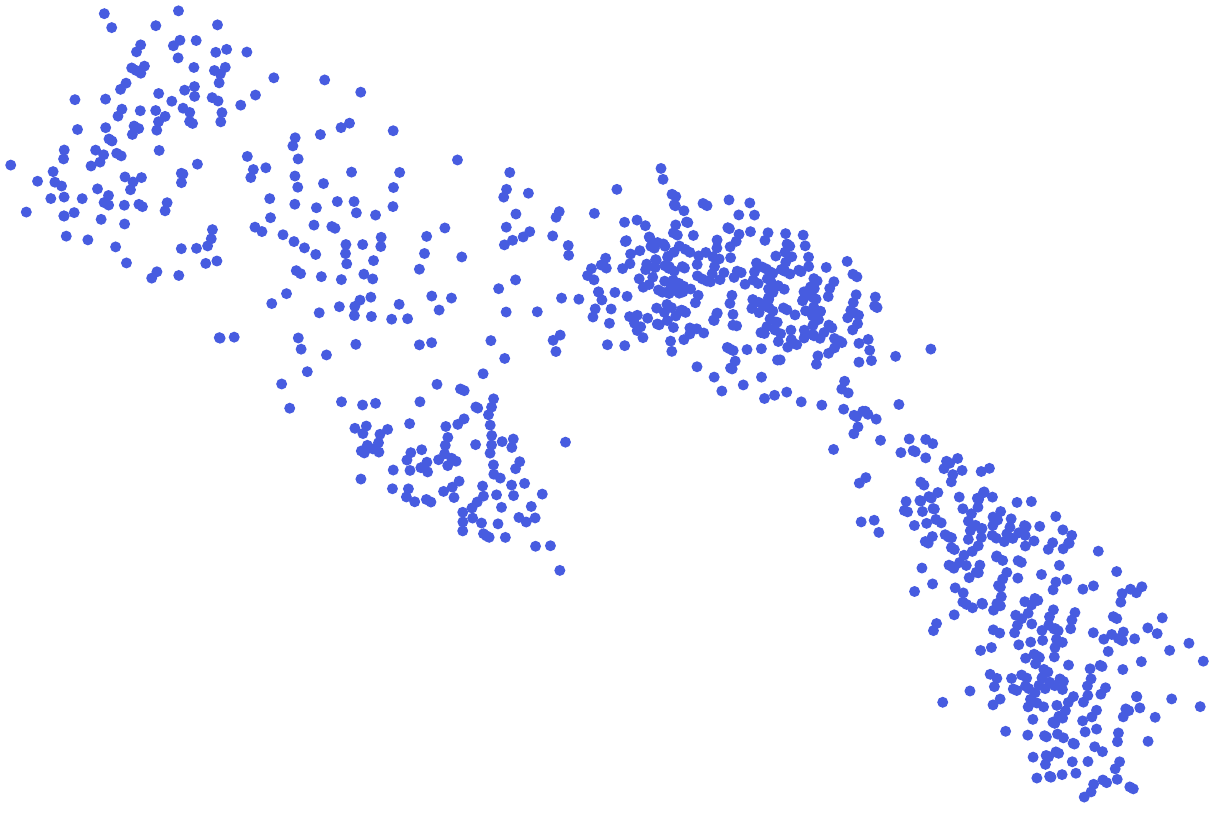}};
\node[inner sep=0pt] (fig2) at (6,0)
    {\includegraphics[width=0.12\textwidth, height=2.3cm]{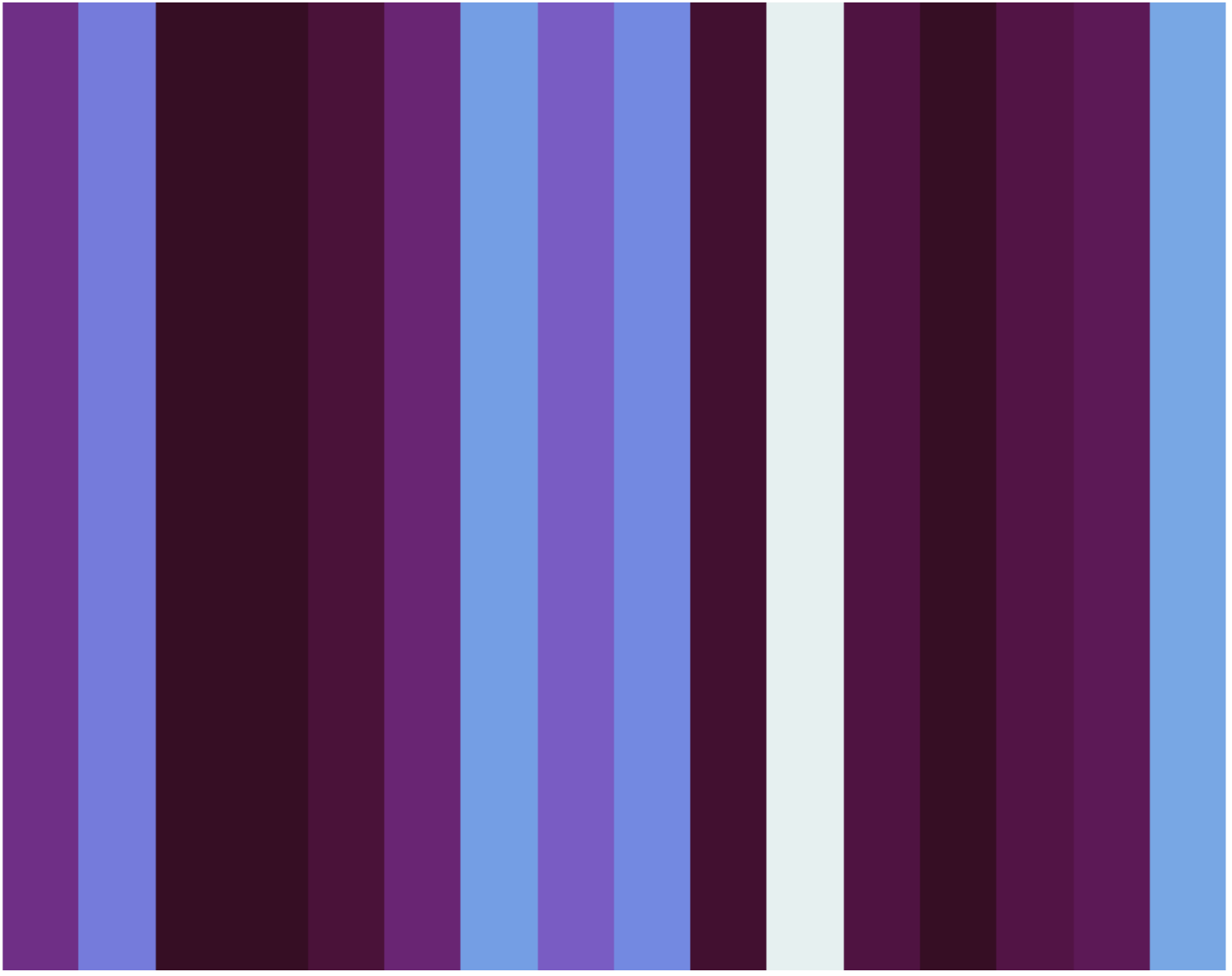}};
    \draw[->, line width=0.2mm] (2.3,0) -- (3.9, 0);
\end{tikzpicture}
\caption{{\it (Top):} non-rotated {\it (Bottom:)} rotated point-cloud with their respective invariant feature responses.}
      \label{fig2}
\end{figure}

{\bf Data augmentation:} As the number of samples ($69$) is much smaller than the number of parameters ($2201$), we use explicit data augmentations as follows: we use uniformly random downsampling scheme to select $512$ points out of $928$ points. This essentially increases the number of training samples significantly, hence the learning is feasible. As claimed before our model is rotation invariant, hence the explicit rotation augmentation does not make sense. In Fig. \ref{fig2}, we show that our model is indeed invariant to rotations.

{\bf Results and Ablation Study:} We achieve $\mathbf{90.72\pm 0.79}\%$ classification accuracy with the sensitivity and specificity to be $87.88\%$ and $94.44\%$ respectively. If we remove the ``attention'' module, we can achieve $72.46\%$ classification accuracy. This clearly indicates the usefulness of the ``attention'' module used in this work.

\section{Conclusions}
Point-cloud helps with understanding 3D geometric shapes for medical data. In this work, we propose an ``augmentation-free" rotation invariant CNN for point-cloud, and apply it on public dataset OASIS for identifying corpus callosum shapes. The core of our method are the proposed rotation invariant convolution on point-cloud by inducing topology from sphere, and the usage of attention mechanism to focus on specific parts of 3D shapes. Our method achieved superior performance with comparatively lean model. In future, we will focus on unified end-to-end trainable model for segmenting ROI from the point-cloud extracted from brain images.
\bibliographystyle{IEEEbib}
\bibliography{references}
\end{document}